\documentclass[12pt]{iopart}

\usepackage{graphicx}
\usepackage{braket}

\newcommand{\refsubfig}[2]{\fref{#1}(#2)} 
\newcommand{\CNOT}{\textsc{cnot}}
\newcommand{\SWAP}{\textsc{swap}}
\newcommand{\Rphi}{R_{\varphi}(\pi/2)} 
\newcommand{\tildeU}[2]{\tilde{U}_{#1}^{(#2)}}

\begin{document}

\title{Universal Control of Ion Qubits in a Scalable Microfabricated Planar Trap}
\author{C D Herold, S D Fallek, J T Merrill, A M Meier, K R Brown, C E Volin and J M Amini}
\address{Georgia Tech Research Institute, Atlanta, GA 30332, USA}
\ead{Creston.Herold@gtri.gatech.edu}
\vspace{10pt}
\begin{indented}
\item[]17 September 2015, updated 19 February 2016
\end{indented}

\begin{abstract}
We demonstrate universal quantum control over chains of ions in a surface-electrode
ion trap, including all the fundamental operations necessary to perform
algorithms in a one-dimensional, nearest-neighbor quantum computing
architecture. 
We realize both single-qubit operations and nearest-neighbor entangling
gates with Raman laser beams, and we interleave the two gate types.
We report average single-qubit gate fidelities as high as 0.970(1) for two-, three-, and
four-ion chains, characterized with randomized benchmarking.
We generate Bell states between the nearest-neighbor pairs of a three-ion chain,
with fidelity up to 0.84(2).
We combine one- and two-qubit gates to perform quantum process tomography of
a \CNOT~gate in a two-ion chain, and we report an overall fidelity of 0.76(3).
\end{abstract}


\vspace{2pc}
\noindent{\it quantum computing, quantum control, trapped ions, surface-electrode trap\/}


\section{\label{set:intro}Introduction}
The size of the Hilbert space describing a quantum computer grows exponentially
with the number of qubits. Despite this, any quantum computation can be
implemented to arbitrary accuracy by composing elements from a set of one- and
two-qubit operations \cite{Barenco1995,Leibfried2003}. Any such operation set is
called a universal gate set. Universal gate sets have been demonstrated with
NMR \cite{Vandersypen2001}, superconducting \cite{Bialczak2010,Chow2012},
neutral atom \cite{Maller2015}, and trapped ion
\cite{Schmidt-Kaler2003a,Hanneke2009} qubits.

Trapped ions are attractive for building an extensible quantum system due to
their long coherence times and ability to be entangled via photons and phonons
\cite{Kim2014,Hucul2014}. Ion trap quantum computer architectures could use
transport \cite{Kielpinski2002} or photonic interconnects \cite{Monroe2013} to
provide connectivity between disparate regions of the quantum computer. These
architectures can benefit from the fabrication precision and scalability of
microfabricated surface-electrode ion traps \cite{Seidelin2006}. While
components of a universal gate set have been shown in microfabricated traps (see
e.g. \cite{Wang2010,Ospelkaus2011,Warring2013,Wilson2014,Mount2015}), a complete universal
gate set has not yet been demonstrated in a single system.

In this work, we demonstrate a universal gate set on two- and three-ion chains
in a surface-electrode trap. Our architecture is based on transporting the chain
sequentially through a single gate location to provide ion addressing; the
gate beams are pulsed on only when the chain is stationary. We perform both
single-qubit operations and nearest-neighbor entangling gates. We measure
single-qubit gate fidelities for chains of two to four ions using randomized
benchmarking, Bell state fidelities in the nearest-neighbor pairs of a three-ion
chain, and process fidelity of a \CNOT~gate in a two-ion chain. The process
fidelity measurement demonstrates our ability to interleave one- and two-qubit
operations in the same experiment. All features are scalable to longer ion
chains. 

\section{\label{sec:Xport}System overview}
\begin{figure}[htb] \centering
		\includegraphics{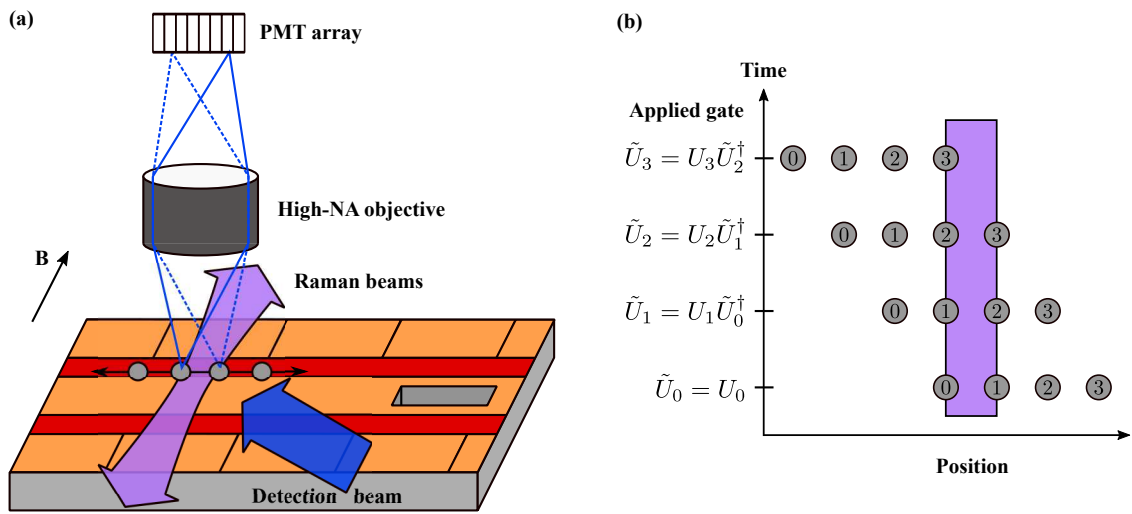}
		\caption{\label{fig:ramanscheme}(a) Schematic diagram (not to scale) of control 
		and individual detection. Both single and two-qubit gates are
		performed with the same Raman gate beams, which address a pair of ions.
		The quantization field $B$ is parallel to the Raman beams.
		(b) With fixed Raman beams, the entire chain is transported
		along the trap axis to address different ion pairs in a ``cascaded''
		fashion. We apply the series of gates $\tilde{U}_i$ to generate
		an arbitrary unitary $U_i$ on each ion $i$.}
\end{figure}

The experimental system is shown schematically in \refsubfig{fig:ramanscheme}{a}.
We trap chains of $^{171}$Yb$^{+}$ ions 60 $\mu$m above the surface-electrode ion trap 
described in \cite{Guise2015}.
This trap features through-chip vias instead of wirebonds. The improved optical
access provided by this feature allows for tight focusing of the Raman gate
beams, which propagate parallel to the surface of the trap.
Qubits are encoded in the hyperfine clock states of $^{171}$Yb$^+$ with 
$\ket{0} \equiv {}^2S_{1/2}\ket{F=0,m_F=0}$ and $\ket{1} \equiv \ket{F=1,m_F=0}$.
Details on trapping, state preparation, and readout of
$^{171}$Yb$^{+}$ ions can be found in \cite{Olmschenk2007a}. 

We load individual ions and merge each ion into a harmonic well. In this way, we
deterministically build chains of a desired number of ions. During state
detection, each ion in the chain is imaged onto a separate channel of a multi-channel
photomultiplier tube (PMT) \cite{Manning2014}. This allows us to detect the
state of each ion in the chain independently. 
Despite efforts to reduce crosstalk by using only every
fourth PMT channel, we observe 5\% crosstalk between ions due
to coma in the imaging system. This crosstalk is calibrated and accounted for
with a joint determination of which ions were bright in a given experiment set.
With 400~$\mu$s of fluorescence collection, our state detection fidelity is 0.98(1).

A single pair of Raman beams provides both one- and two-qubit gates. The beams
are derived from a mode-locked tripled YAG laser, in a configuration similar to
that described in \cite{Islam2014,Hayes2010}. The circularly polarized beams
propagate perpendicular to the trap axis and parallel to the quantizing magnetic
field. In this configuration, the gate beams
do not couple to the axial motional modes or the $\ket{F=1,m_F=\pm1}$ Zeeman levels.
Furthermore, the trapping potentials are designed so that one set of radial
secular modes is rotated normal to the beam direction. This limits coupling of
the Raman beams to only a single
set of motional sideband transitions along with the carrier,
reducing the effect of off-resonant shifts 
and removing the dependence of the carrier transition on the temperature of these modes. 
After Doppler cooling all modes, we perform sideband cooling with the Raman
beams only on the radial mode used for gates, resulting in sub-quanta occupation
of excited motional states.

\begin{figure}[htb]
	\centering
	\includegraphics[width=0.75\textwidth]{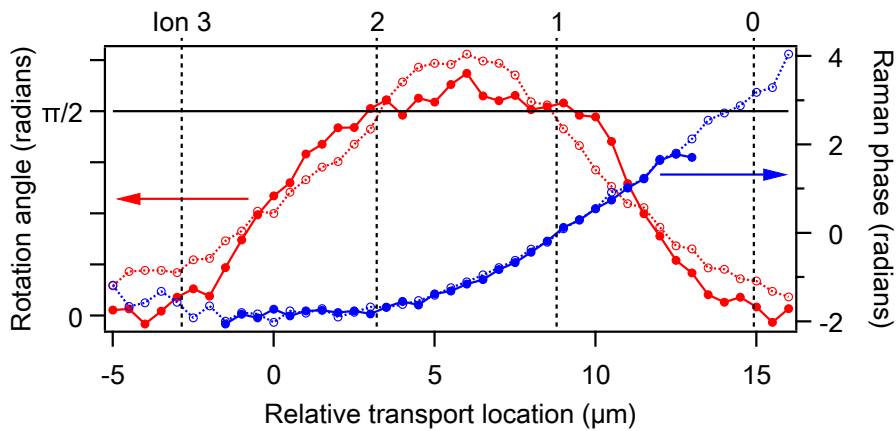}
	\caption{\label{fig:crosstalk}Characterization of the intensity profile and
	phase front of the Raman control beams. The position of each ion in a 4-ion
	chain is shown across the top when ions 1 and 2 are addressed. The beam
	power was calibrated for $\pi/2$ rotations on the addressed pair. Open
	(filled) circles use bare (PB1 stabilized) pulses, and markers are connected
	for clarity. The PB1 pulse sequence stabilizes Rabi rate errors, resulting
	in the flatter slope at rotation angle $\pi/2$ around ions 1 and 2.
	Additionally, for bare rotation angles
	less than $\pi/4$, the PB1 sequence suppresses rotations.
	The slight asymmetry between ion 0 and 3 is due to beam coma.}
\end{figure}

M{\o}lmer-S{\o}rensen (MS) gates between pairs of neighboring ions
provide the entangling gates for this system. The Raman beam in one direction
(consisting of the red- and
blue-detuned beams\footnote{These beams are generated with a single acousto-optic
modulator, the output of which is imaged onto the trapping region with several
telescopes to ensure the two beams overlap at their foci.}) is focused to a waist
($1/e^2$ half-width of the intensity profile) 
of roughly 7 $\mu$m while the opposing
Raman beam has a waist of 13.5 $\mu$m. The average power in these beams is on the
order of 10 and 100 mW, respectively. The $1/e$ half-width of the Raman interaction
Rabi rate (see \fref{fig:crosstalk}) is approximately equal to our ion spacing of
about 6~$\mu$m, permitting pairwise addressing of ions within a chain as in
\fref{fig:ramanscheme}. The Raman beams remain static throughout the course of
the experiment; we transport the ion chain to bring different pairs of ions into
the beam.

Still addressing ions pairwise,
we employ the cascading transport scheme shown in \refsubfig{fig:ramanscheme}{b}
to implement an arbitrary set of one-qubit unitary operations $\{U_i\}$, where
$i$ is the ion index. The ion chain is first transported such that the
$i=0,1$ ion pair (i.e. the 01 pair) is illuminated by the laser.
The unitary $\tilde{U}_0=U_0$ is performed,
leaving the endmost ion in the desired state. Then the ions are transported to 
address the subsequent pair. 
The operation $\tilde{U}_1 = U_1 \tilde{U}_0^{\dagger}$ 
results in the net operation 
$\tilde{U}_1 \tilde{U}_0 = U_1 U_0^{\dagger} U_0 = U_1$
on the second ion, accounting for the effect of the first operation. 
This procedure is repeated sequentially along the chain until all ions
are targeted, resulting in the desired gate on each ion.
Universal control is achieved in this transport architecture by interleaving
such cascaded single-qubit gates with pairwise MS gates.

In this work, ions are only transported over very short distances
corresponding to the spacing between ions or the length of the chain. Each
transport step is performed adiabatically in 100 $\mu$s. We varied the transport
duration between 100 $\mu$s and 1 ms but could not see any effect on gate
fidelity at the present error levels due to heating.
In addition to shuttling the ions through the gate beams, we adjust the 
strength of the axial harmonic well to optimize the ion spacing for each 
type of gate. For two-qubit operations, where a strong coupling between 
the ions' motion is desired, the axial well is stronger than for single qubit operations, 
where crosstalk to unaddressed ions needs to be minimized. 
For Doppler and sideband cooling, the typical axial trap frequency is $2\pi
\times 0.5$ MHz, and the radial frequencies are $2\pi \times (1.8, 2.1)$ MHz.
The lower radial frequency corresponds to the mode addressed for all gates.
We reduce the axial frequency by a factor of $\sqrt{2}$ from the cooling trap
frequencies to spread the ions out for single-qubit gates, and we increase it by
$\sqrt{1.5}$ times to further split the radial modes for two-qubit operations.
Similarly, for detection the ion spacing is optimized to match
the spacing of the photomultiplier array.

\section{\label{sec:sq}Single-Qubit Control}
As described above, all single-qubit control is implemented via pairwise
addressing. 
For each ion pair, we calibrate the ion chain position to balance the carrier
transition Rabi rate on the two addressed ions. We then calibrate the gate beam
intensity to give a $\pi/2$ rotation in 4~$\mu$s.
These calibrations are stabilized against Rabi rate errors using PB1 compensating
pulse sequences \cite{Wimperis1994,Brown2004,Brown2005}, which are composed of
the calibrated $\pi/2$ rotations. Including hardware programming overhead, a PB1
composite $\pi/2$ operation is performed in 120 $\mu$s. 
Arbitrary single-qubit unitaries are
constructed from these composite $\pi/2$ rotations about different axes in the
equatorial plane of the Bloch sphere. The angle the axis of rotation makes with
the $x$-axis is set by the phase of the Raman beams and is controlled via an
acousto-optic modulator. We developed a compiler (see \ref{app:compiler}) to
efficiently implement arbitrary unitaries in the fewest $\pi/2$ rotations.

\begin{figure}[htb]
	\centering
	\includegraphics[width=0.85\textwidth]{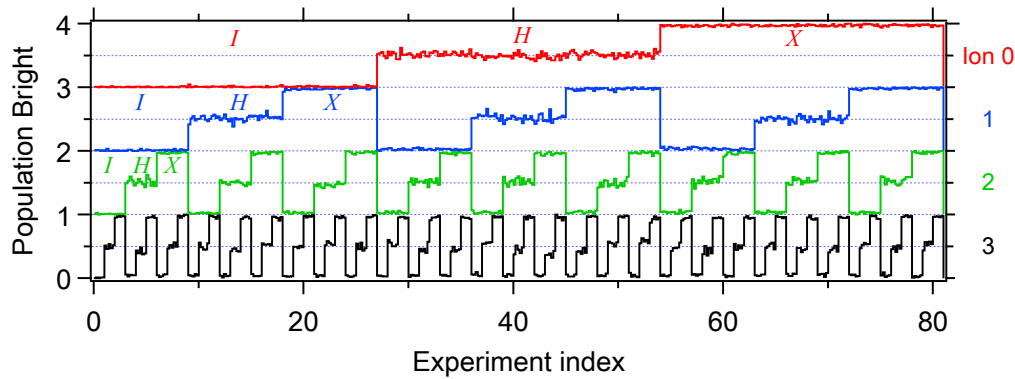}
	\caption{\label{fig:prep}As a demonstration of independent single-qubit 
	control through pairwise addressing, we perform all combinations of the set
	$\{I,H,X\}$ on each qubit in a 4-ion chain after initializing to $\ket{0000}$. 
	Probabilities are offset one unit for each ion. 
	Each unique preparation is labeled by an experiment index and was repeated 
	five times to show the spread in results.}
\end{figure}

Implementation of an arbitrary set of unitaries across the chain under actual
experimental conditions is more complicated than the idealized scheme presented
in \refsubfig{fig:ramanscheme}{b}. First, wavefront curvature across the gate
beams causes ions to experience a position-dependent phase of the Raman
interaction (see \fref{fig:crosstalk}), resulting in a spatially varying
$\pi/2$ rotation axis. Second, the pairwise addressing of our gate beams is
imperfect, and ions adjacent to the addressed pair are observed to flop at
reduced but significant rates. To a large degree, this crosstalk is mitigated
by our choice of compensating pulse sequence since PB1, a so-called ``passband''
sequence, also suppresses small excitations (\fref{fig:crosstalk}). Due to
comatic aberrations, the crosstalk is asymmetric, and it is a reasonable
approximation to treat the right-hand side of our gate beam as crosstalk-free
when PB1 is used. Therefore, we only calibrate and correct for the small rotation
angle of left-side crosstalk. This asymmetry dictates the direction in which we
cascade single-qubit operations. As an example of independent single-qubit
control, we prepare each ion in a four ion chain in all $3^4$ combinations of
$\{\ket{0},\ket{1},\ket{+}=\left(\ket{0}+\ket{1}\right)/\sqrt{2}\}$
(\fref{fig:prep}).

\begin{figure}[htb]
	\centering
	\includegraphics[width=0.85\textwidth]{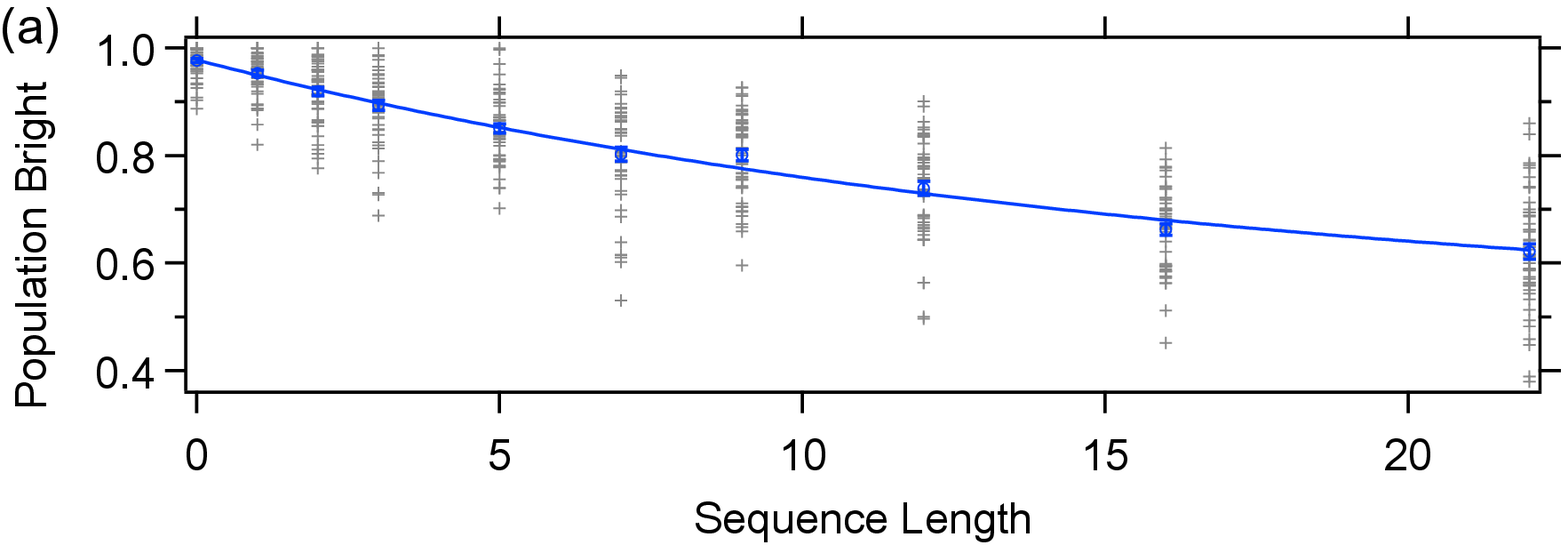}
	\includegraphics[width=0.85\textwidth]{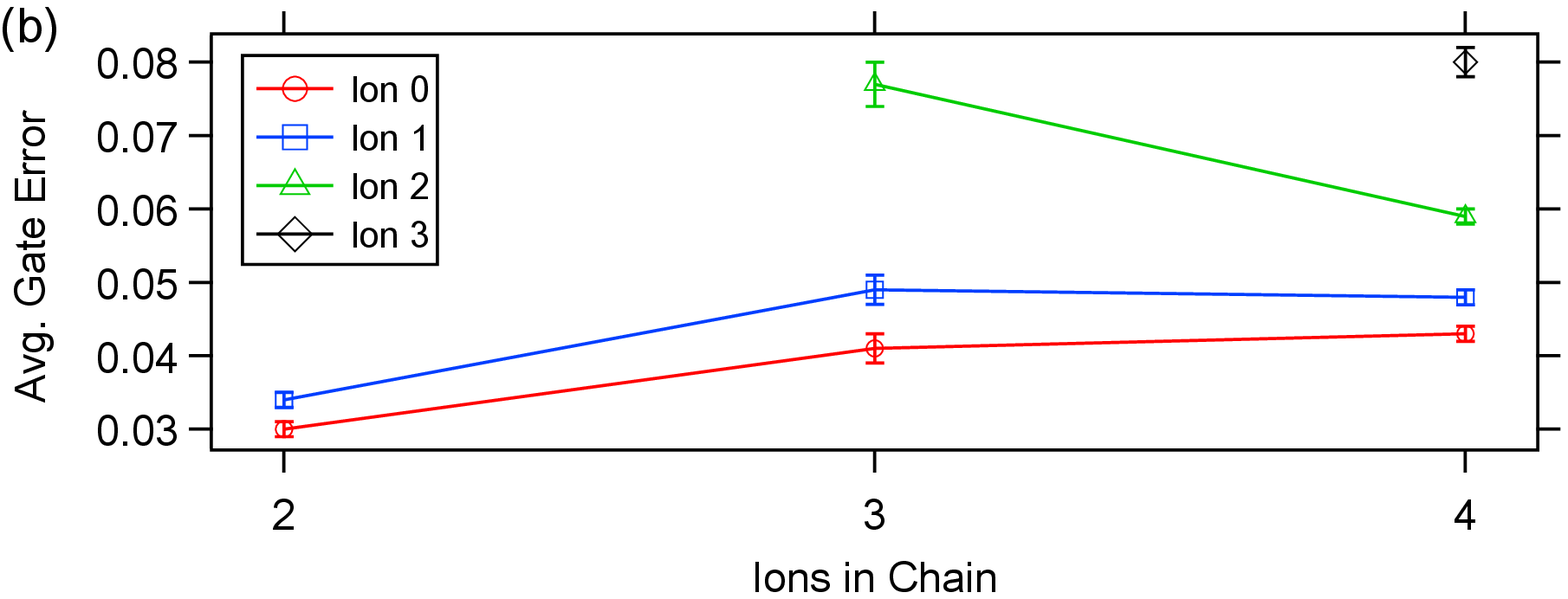}
	\caption{\label{fig:rb}(a) Typical randomized benchmarking data, here for
	ion 1 in a 2-ion chain. Gray crosses show the data spread at each
	sequence length, and the blue circles are the average data with error bars
	reflecting the standard error in the mean. The fit is described in the text.
	(b) Average Clifford gate error for each ion as a function of ion chain
	length. Error bars represent the statistical error of the fit.}
\end{figure}

To characterize the single-qubit gate fidelity, we employ randomized
benchmarking \cite{Knill2008}. Starting with all ions in state $\ket{0}$, a
randomly chosen Clifford gate is performed on each ion, cascading the gates
across the chain as described above. Each Clifford operation requires on average
400 $\mu$s of Raman pulses and is followed by 100 $\mu$s transport between gate
locations.  Subsequent passes build a random sequence
of Clifford gates of length $L$ on each ion. The final pass consists of an inverting
gate calculated to take each ion to $\ket{1}$. The results for one ion in a
two-ion chain are shown in \refsubfig{fig:rb}{a} for varying sequence lengths. 
We performed benchmarking on two- to four-ion chains and fit the average gate
fidelity $\bar{F}_{\mathrm{RB}}$ for each ion, following \cite{Brown2011}, to
\begin{equation}
		\bar{F}_{\mathrm{RB}} = \frac{1}{2} + \frac{1}{2}
		\left(1 - 2\varepsilon_m\right)
		\left(1 - 2\varepsilon_g\right)^L,
		\label{eq:RBfit}
\end{equation}
where $\varepsilon_g$ is the average gate error and $\varepsilon_m$ includes
error from state preparation and measurement as well as error in the inverting
gate. The average $\varepsilon_m$ for fits on all chain lengths is 0.025(2).
We also performed a three parameter fit as
in \cite{Magesan2012}, where the constant offset is allowed to vary from
$\frac{1}{2}$, however these fits gave non-physical results with negative
$\varepsilon_m$. 
Because the measurement imbalance in our apparatus is small,
the two-parameter fit with offset fixed at $\frac{1}{2}$ is justified; a typical
reduced-$\chi^2$ for the fits is 1.6.

As can be seen in \refsubfig{fig:rb}{b},
we measure average single-qubit gate fidelity as high as 0.970(1) for two-ion
chains. However, the average gate error
increases with ion chain length and position within the chain in the cascade
direction. 
Longer chains require more calibrations and more operations, which lead to more
time for dephasing between successive operations on a given qubit. As a result,
one expects gate error to increase with chain length. 

That the end ions, i.e.~ions 2 and 3, have a higher gate error can be explained
by imperfect crosstalk correction. As we noted in the discussion of
\fref{fig:crosstalk}, the wings of our comatic gate beams cause small-rotation
crosstalk on ions adjacent to the addressed pair. The PB1 sequence suppresses
the average value of this crosstalk to the point where it is negligible on the
left-hand (low ion number) side of the beam; on the right-hand side, we
calibrate and correct for the remaining crosstalk in our cascaded gate
sequences. Therefore, ions 0 and 1, unlike ions 2 and 3, are not subject to this
crosstalk. Despite our careful calibration, we suspect that the increased slope
due to PB1 (\fref{fig:crosstalk}) may be \textit{increasing} our sensitivity to
crosstalk fluctuations on the right-hand (high ion number) side of the beam.
Improving the passive stability of our gate beams should increase the overall
single-qubit gate fidelity. We note that the average gate error
for ion 2 decreased from a three-ion to four-ion chain.
This is most likely due to relatively poorer calibration; systematic
errors are not included in the error bars.

\section{\label{sec:2q}Two-Qubit Gates}
\subsection{\label{sec:only2}Two-Ion Entanglement}
\begin{figure}[ht]
	\centering
	\includegraphics[width=0.8\textwidth]{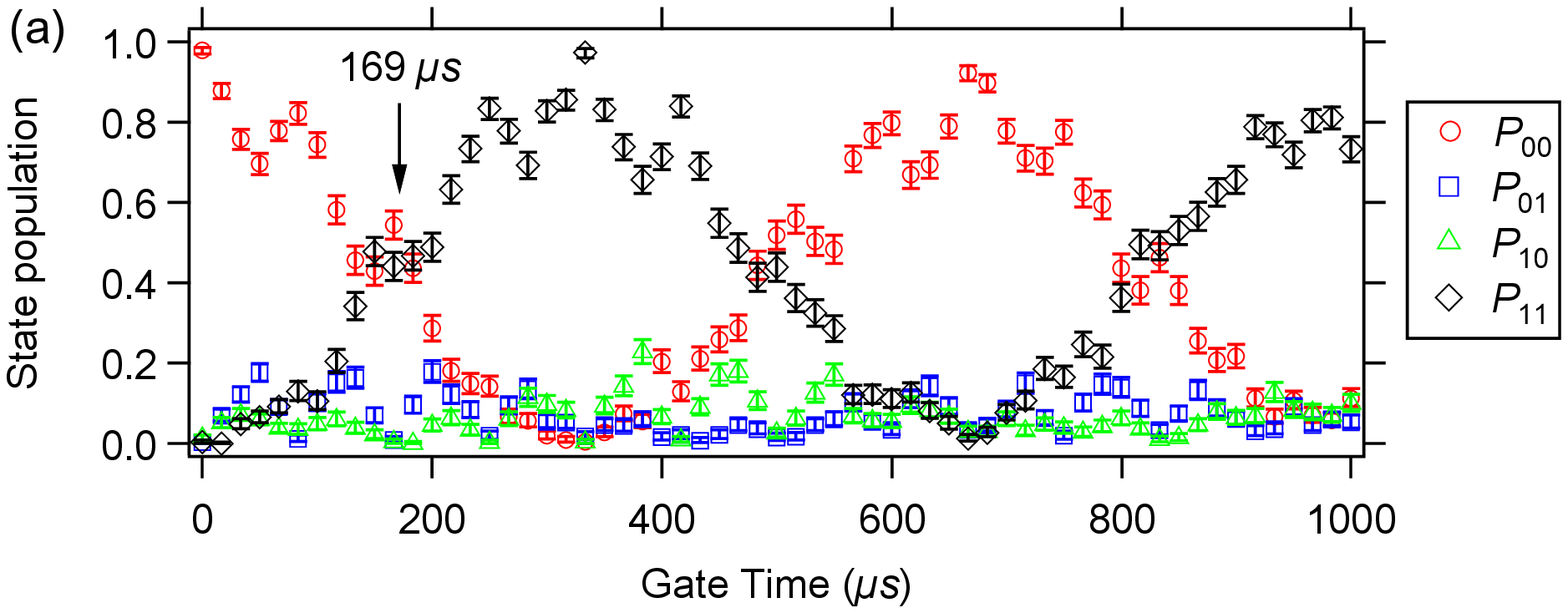}
	\includegraphics[width=0.8\textwidth]{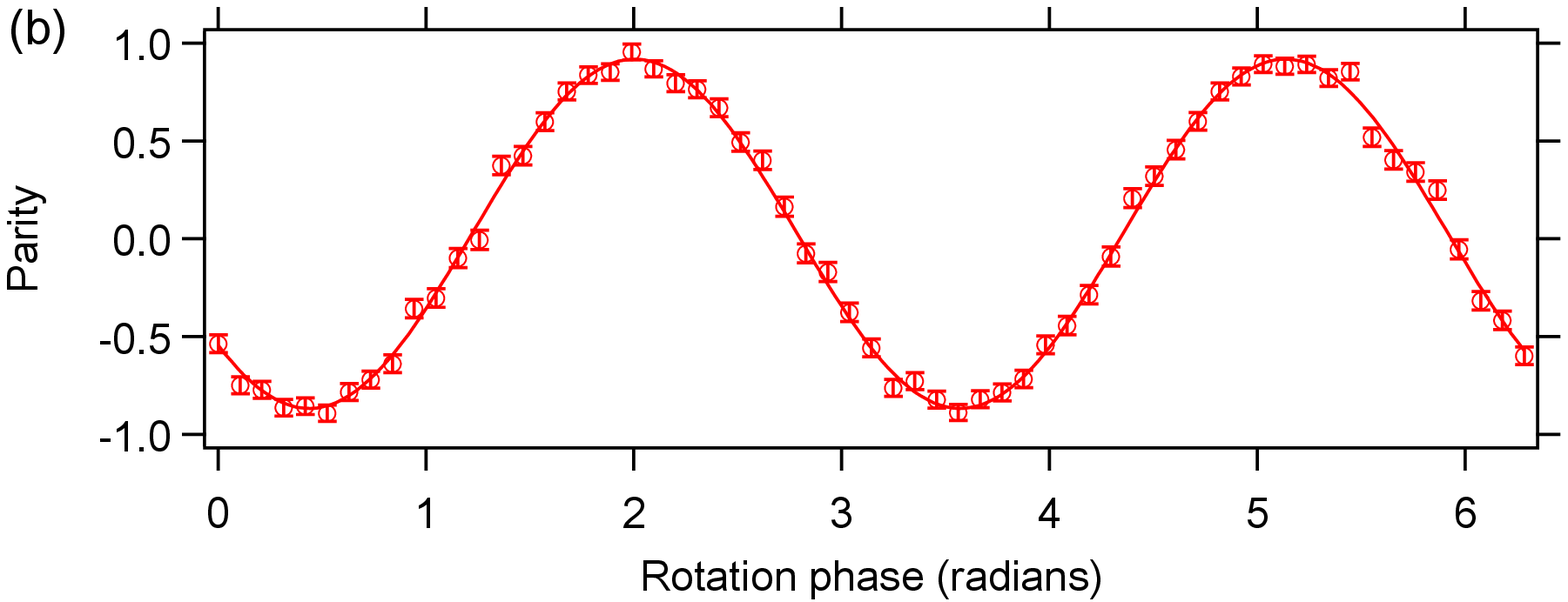}
	\caption{\label{fig:MS} (a) Population evolution during the
	M{\o}lmer-S{\o}rensen interaction of the four two-ion states: $P_{00}$,
	$P_{01}$, $P_{10}$, $P_{11}$.
	(b) Parity measurement $(P_{00}+P_{11}-P_{01}-P_{10})$ of the 
	entangled ions at gate time of $169~\mu$s.}
\end{figure}

To complete our universal gate set, we require an entangling gate.
Here we employ the M{\o}lmer-S{\o}rensen (MS) gate on a pair of ions \cite{Sørensen2000}.
During MS gates, we strengthen our axial confinement to increase the frequency
splitting of our radial modes in comparison to single-qubit gates.
This reduces motional excitation of undesired modes, yielding higher fidelity. 
State evolution of a typical gate is shown in \refsubfig{fig:MS}{a}
where ion motion has been excited and de-excited twice per gate,
i.e. two ``loops'' per gate, resulting in a 169 $\mu$s gate time. This is
substantially faster than our PB1 compensated implementation of single-qubit
unitaries.

To verify entanglement following the MS gate, Bell state fidelity estimation is
performed using the ``parity flopping'' technique \cite{Sackett2000}.
The resulting parity curve is shown in \refsubfig{fig:MS}{b}. 
Bell state fidelity is measured to be $F_{\mathrm{Bell}}= 0.93(2)$.
We expect this fidelity is largely limited by imperfect overlap of the red and blue
sideband beams during the MS interaction; a redesigned optics system will remedy this
issue. This MS gate, along with the individual qubit rotations 
in \sref{sec:sq}, comprise our universal gate set for a two-ion chain. In
\sref{sec:CNOT}, we demonstrate combined one- and two-qubit gates to
produce a \CNOT~gate.

\subsection{\label{sec:2of3}Pairwise Entangling Gates on Three Ions}
Universal control of ion chains requires the ability to entangle any pair
of ions. Within our architecture, we entangle neighboring ions in pairs.
However, this entanglement can be transferred to non-neighboring pairs through
\SWAP~gates, which can be composed from \CNOT~gates. 
In this section, we characterize entangling gates on pairs of neighboring ions within a
three-ion chain, the last component of a universal gate set for three ions.

To estimate the MS gate fidelity, we measure $F_{\mathrm{Bell}}$ as in the
previous section.
All ions are prepared in $\ket{0}$, the two targeted ions are
shifted into the laser beam, then an MS gate and analysis operation are performed.
Parity is calculated using only the state populations where the unaddressed ion
remains in the desired $\ket{0}$ state, and the resulting fidelities are summarized
for each targeted pair in \tref{tab:2of3}.

\begin{table}
	\caption{\label{tab:2of3}Parity measurement for pairwise entanglement in a three-ion 
	chain. $P_{00}$ and $P_{11}$ are the measured populations for the state
	where the unaddressed ion is in $\ket{0}$, i.e. for an MS gate on the 12-pair, the
	states $\ket{000}$ and $\ket{011}$. Similarly, the parity amplitude is
	extracted with the unaddressed ion in state $\ket{0}$. 
	$P_1$ is a measure of the population build-up on
	the unaddressed ion, where we trace over the addressed ions.}
 	\begin{indented}
	\lineup
	\item[]\begin{tabular}{@{}rccccc}
		\br
 		Ion pair & $P_{00}$ & $P_{11}$ & Parity amp. & $P_{1}$ & $F_{\mathrm{Bell}}$ \\ 
 		\mr
 		No echo 01 & 0.32(1) & 0.45(1) & 0.66(1) & 0.18(1)\0 & 0.72(1) \\
 		Echo 01 & 0.37(1) & 0.46(1) & 0.62(1) & 0.10(1)\0 & 0.72(2)\\
 		\mr
 		No echo 12 & 0.47(1) & 0.41(1) & 0.81(1) & 0.094(7) & 0.84(2) \\
 		Echo 12 & 0.41(1) & 0.47(1) & 0.76(1) & 0.023(3) & 0.82(2) \\
		\br
 	\end{tabular}
 	\end{indented}
\end{table}

We observe moderate build-up of undesired population in the $\ket{1}$ state on the
unaddressed ion during the MS gate; this is reported as $P_1$ in
\tref{tab:2of3}. Due to the spatial asymmetry of our gate beam (shown in the
rotation angle of \fref{fig:crosstalk}), the MS crosstalk on the 01 pair is
worse than the 12 pair. In \ref{app:echo}, we describe an ``echoing''
sequence which decouples a subset of ions from a general MS interaction. For an
MS gate on two ions in a three-ion chain, we perform the MS gate with two
motional loops. The echo is inserted after each loop and flips the spin of the
untargeted ion (or equivalently the spins of the targeted ions) with a $Y$-gate.
Early results show that the echo sequence reduces $P_1$; however, the Bell-state
fidelity is not improved. This is likely due to the error introduced by the
additional single-qubit operations used in the echo. We expect that increased
single-qubit gate fidelity should make the echo more effective and allow us to
perform high-fidelity MS gates on any adjacent pair of ions in longer chains.

\section{\label{sec:CNOT}CNOT Characterization}
\begin{figure}[h] \centering
	\includegraphics[width=0.8\textwidth]{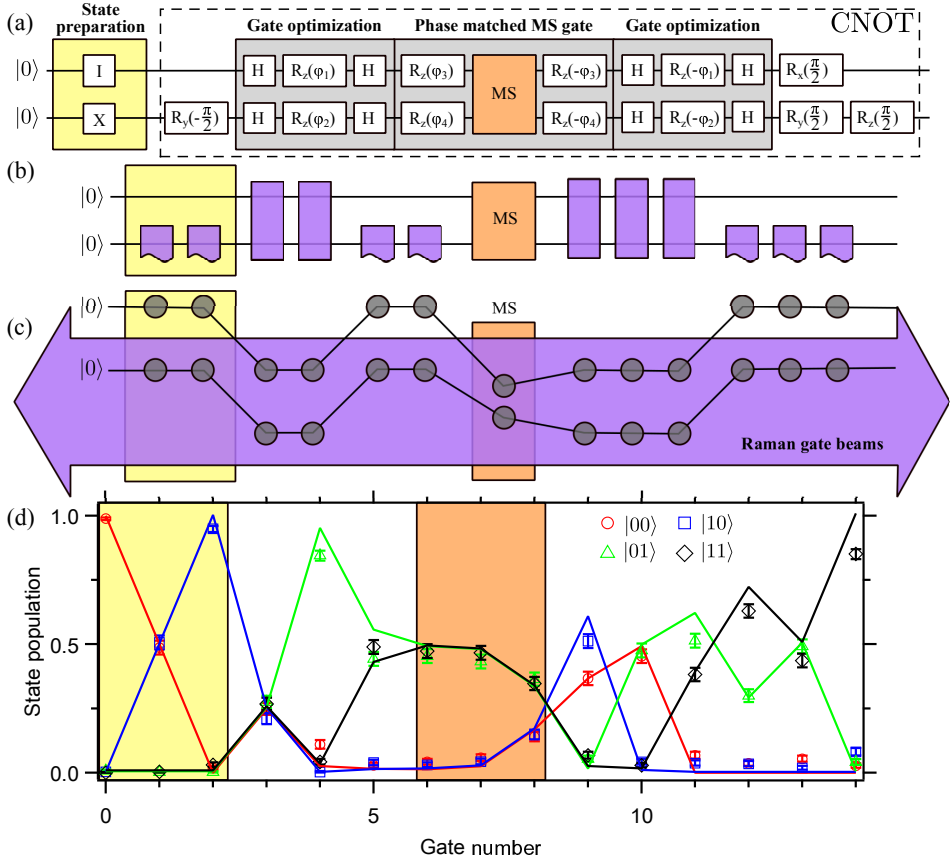}
	\caption{\label{fig:CNOTseq}
	(a)~Circuit diagram to build a \CNOT~gate (dashed box, controlled by lower
	ion) from single-qubit
	rotations and an MS gate (orange box). After initialization to $\ket{00}$,
	the ions are prepared in $\ket{10}$ (yellow box).
	(b)~Schematic of pairwise addressing for compiled \CNOT. Each purple block
	represents a PB1 pulse sequence stabilized $\pi/2$ rotation. 
	(c)~Ion transport during \CNOT~implementation. 
	(d)~Measured population evolution throughout our compiled \CNOT. Each
	numbered gate is a $\pi/2$ rotation
	except for the orange region in which the MS gate is shown in 2 steps. Simulated
	expectations are connected with solid lines and agree very well with the
	data. For input state $\ket{10}$, the \CNOT~gate yields the expected output state
	$\ket{11}$ with probability 0.85.}
\end{figure}

For a demonstration of a small circuit using
our universal gate set, we perform the \CNOT~gate on two ions. 
The \CNOT~is built from an MS gate wrapped with a particular set of single qubit rotations, 
as shown in \refsubfig{fig:CNOTseq}{a}. 
In \refsubfig{fig:CNOTseq}{b,c} we represent how the \CNOT~circuit fits into the 
cascading architecture discussed in \sref{sec:Xport}. Single-qubit rotations
are performed in two cascading passes, one before the MS gate and one after. As in
\sref{sec:2q}, axial confinement is increased for the MS gate and then
subsequently relaxed for single-qubit operations, adding an additional 50 $\mu$s
for each. 
We utilize four steps of transport (100 $\mu$s each, as before), 
excluding the transport needed for state preparation and detection.

Successful combination of one- and two-qubit operations requires accounting for
the Raman phase difference due to increased axial confinement during MS gates.
We add $z$-rotations ($R_{z}$) before and after the MS gate (with rotation
angles $\varphi_3$ and $\varphi_4$) to ensure it is in phase with the
single-qubit rotations. 
In addition, we add the optimization operations shown in \refsubfig{fig:CNOTseq}{a}. 
These gates do not affect the outcome of the ideal circuit, but they allow us to 
minimize the number of physical operations in the compiled single-qubit cascades
by varying $\varphi_1$ and $\varphi_2$.
Further discussion of this procedure is provided in \ref{app:compiler}.

As a diagnostic tool, we measure the population evolution throughout the
implementation of the \CNOT~(markers in \refsubfig{fig:CNOTseq}{d}). 
To measure the populations at a given point within
the \CNOT~sequence, we trigger an override on the gate beams, shutting down all
subsequent operations. This freezes the populations, which are read out at the
end of the \CNOT~sequence.
When combined with on-the-fly simulation of ideal
rotations (points connected by solid lines), such a scan has proved a useful
tool for diagnosing calibration errors.
We have also built an optimization routine into our simulation tool to find the
best parameter calibrations to match observed populations. Thus, from a small
set of carefully chosen gate scans like \refsubfig{fig:CNOTseq}{d}, we can
quickly improve overall fidelity.

\begin{figure}[htb] \centering
		\includegraphics[width=1\textwidth]{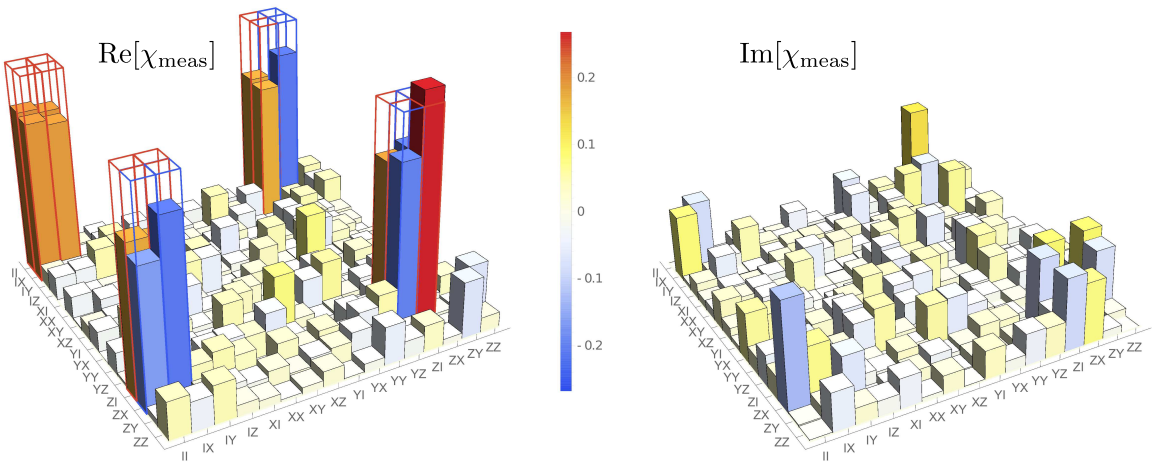}
		\caption{\label{fig:CNOTQPT}Quantum process tomography of \CNOT~on a
		2-ion chain. Bar height reflects magnitude of the real (imaginary) part
		in the left (right) plot with color giving the sign. Ideal matrix values
		are shown as colored wire frames. All imaginary elements should be
        identically zero. The apparent \CNOT~fidelity from QPT is 0.76(3).
		Figure data included as supplementary data on-line.}
\end{figure}

To systematically characterize our \CNOT~gate, we perform quantum process
tomography (QPT) \cite{Nielsen2010}. We prepare and analyze each ion in all
combinations of the set
$\{\ket{0},\ket{1},\ket{+},\ket{-}_y=\left(\ket{0}-\rmi\ket{1}\right)/\sqrt{2}
\}$ resulting in 256 measurements. The preparation and analysis gates are 
\emph{not} compiled into the single-qubit gates in the \CNOT~itself. Rather, two
additional passes of single-qubit operations are used, similar to the state
preparation (yellow box) in \fref{fig:CNOTseq}. Each measurement is repeated
160 times and the maximum likelihood probability to be in $\ket{00}$ is used to
generate the measured process matrix $\chi_{\mathrm{meas}}$, shown in
\fref{fig:CNOTQPT}. The fidelity of our operation with the ideal \CNOT~process,
$\chi_{\mathrm{ideal}}$ is measured as $F = \mathrm{Tr}(\chi_{\mathrm{ideal}}
\chi_{\mathrm{meas}})= 0.76(3)$, where we have reported the mean and standard
deviation of $10^4$ parametric bootstrap resamplings of our measurements. 
From the resampled $\chi_{\mathrm{meas}}$, we estimate a typical matrix element
error bar to be 0.02. There are only 16 nonzero matrix elements of
$\chi_{\mathrm{ideal}}$, all in the real part. Nearly all other measured matrix
elements are consistent with zero within three standard deviations.

Using a simple uncorrelated error model to account for four single-qubit gates
(from randomized benchmarking fidelity) and one MS gate (from Bell state fidelity),
we expect an approximate fidelity of
$(\bar{F}_{\mathrm{RB}})^4 F_{\mathrm{Bell}} = (0.97)^4 \times 0.93 = 0.82$. 
Yet, single-qubit errors in preparation and analysis also reduce QPT fidelity
estimates. With the same QPT procedure, we measure a two-qubit
identity operation fidelity of 0.95, leading us to estimate a QPT
fidelity of 0.78 in good agreement with our measurement. 

\section{\label{sec:out}Outlook}
We have demonstrated a universal gate set for two- and three-ion chains for the
first time in a surface-electrode ion trap. Universality is realized through a
novel control scheme which leverages the precision transport capability of our
trap. This scheme is extensible to longer chains. Importantly, we demonstrated
universal control in a \emph{single} experiment by implementing a circuit
consisting of single-qubit rotations and an MS gate to perform a \CNOT~gate.
Single-qubit gates on four-ion chains were characterized as well as echoed MS
gates on pairs within a three-ion chain. Minor system improvements currently
underway should improve fidelities to levels where basic quantum algorithms are
within reach.

Our results bolster the promise of scalable ion trap quantum computing. Through
the integration of junctions, microfabricated ion traps have already been
extended to two-dimensional architectures. Integrated optics are a demonstrated
path to a scalable photonic interface with ions in large trap arrays. We have
shown that microfabricated traps are also compatible with high-fidelity
universal quantum control of ion chains.
 
\ack 
This material is based upon work supported by the Office of the Director of
National Intelligence (ODNI), Intelligence Advanced Research Projects Activity
(IARPA) under U.S. Army Research Office (ARO) contract W911NF1010231. All
statements of fact, opinion, or conclusions contained herein are those of the
authors and should not be construed as representing the official views or
policies of IARPA, the ODNI, or the U.S. Government.

\appendix

\section{\label{app:compiler}Compiler}
In our control scheme, we must implement arbitrary unitary operations on individual 
qubits for universal control. Even for algorithms designed around a discrete set
of single-qubit gates, accounting for the crosstalk shown in
\fref{fig:crosstalk} would require unitary operations outside that set. Here,
we describe how we construct arbitrary unitaries from our calibrated,
stabilized, physical operations. Additionally, methods to minimize the number of
physical rotations required are detailed.

As discussed in \sref{sec:sq}, by adjusting the phase of our Raman gate laser,
we can perform rotations about any axis in the equatorial plane of the Bloch
sphere. Defining the angle that axis makes with the $x$-axis as $\varphi$, our
physical operations $\Rphi$ are composite $\pi/2$ rotations stabilized with the
PB1 sequence. While we have constructed various analytic solutions to perform
arbitrary unitaries in a fixed number of axis angles $\varphi_j$, direct
numerical search has proved more efficient. In a test of 1000 randomly generated
unitaries, we always find a solution in three or four rotations, with an average
of 3.25.

Our compiler takes a set of requested circuit-level gates for the ion chain and
computes a target unitary $U_i$ for each ion $i$. First, $U_0$ is performed on
the 01 ion pair, using the rotations calculated from our numerical search. While
this leaves ion 0 in the desired state, the physical operations act in a
$z$-rotated basis due to the phase front curvature of the Raman gate beams
(\fref{fig:crosstalk}). Each physical rotation on ion 1 is also through an
angle of $\pi/2$; the rotation due to the crosstalk on ion 2 is much
smaller. Neglecting crosstalk beyond next-nearest neighbors, the resulting
physical operations are summarized in step 1 of \tref{tab:sqcross}, where
the net physical unitary $\tildeU{i}{j}$ denotes the physical effect on ion
$j$ of a unitary targeting ion $i$. In each step, the unitary $\tildeU{i}{j=i}$ is
performed, while the compiler
stores a unitary corresponding to the calibrated crosstalk on $j=[i+1,i+2]$.
In step 2, we then need to perform the product of the targeted
unitary $U_1$ and the inverse of the stored unitary from the previous step. Thus,
in each subsequent step, we numerically search for physical rotations whose
net effect is to undo the crosstalk from previous operations and perform the
targeted unitary $U_i$.

\fulltable{\label{tab:sqcross}In contrast to the idealized pairwise control
	shown in \refsubfig{fig:ramanscheme}{b}, here we show how our compiler 
	must account for crosstalk.}
		\br
		& Addressed & & & & \\
		\ns
		Step & Pair & Ion 0 & Ion 1 & Ion 2 & Ion 3 \\
		\mr
		1 & 01 & $\tildeU{0}{0} = U_0$ & $\tildeU{0}{1}$ & $\tildeU{0}{2}$ & -- \\
		2 & 12 & -- & $\tildeU{1}{1} = U_1 \left(\tildeU{0}{1}\right)^{\dagger}$
		& $\tildeU{1}{2}$ & $\tildeU{1}{3}$ \\
		3 & 23 & -- & -- 
		& $\tildeU{2}{2} = U_2 \left(\tildeU{1}{2} \tildeU{0}{2}\right)^{\dagger}$
		& $\tildeU{2}{3}$ \\
		4 & 34 & -- & -- & --
		& $\tildeU{3}{3} = U_3 \left(\tildeU{2}{3} \tildeU{1}{3}\right)^{\dagger}$ \\
		\mr
		\centre{2}{Net gate:} & $U_0$ & $U_1$ & $U_2$ & $U_3$ \\
		\br
\endfulltable

To correct crosstalk, we store a unitary corresponding to the history for each
ion throughout the algorithm. However, we can take further advantage of
this storage to reduce the number of applied physical $\Rphi$, thereby
improving overall fidelity. This is implemented in two ways. 
First, subsequent single-qubit gates are concatenated for each ion, whenever
possible. If they do not need to be performed, we append the inverse operation
to the stored unitaries. At the next physical operation cascade, the inverse
history is composed with any additional requested gate, allowing us to perform
multiple single qubit gates and crosstalk corrections in a single cascade of
operations. Second, we often perform requested unitaries up to a stored $R_z$.
An arbitrary unitary can almost always be performed in only two $\Rphi$
rotations, up to a missing $R_z$, which is nearly half that required for an
exact unitary. This has the effect of transforming an ion's relative frame and
can simply be appended to the stored unitary. Subsequent operations on this ion
need only shift the Raman gate phase to account for the frame shift. In fact,
without global operations, there is no need for a common definition of zero
phase. Instead we define a phase zero for each ion by the Raman beam phase front
when that ion is targeted.

In some instances, performing unitaries up to a phase still produces an exact
result. For example, state preparation on the fiducial state with every ion in
$\ket{0}$ is the same with an $R_z$ inserted:
\begin{equation}
		U_{\mathrm{exact}}\ket{0} = \left(U_{\mathrm{exact}}
		R_z^{\dagger}\right) R_z \ket{0} 
		= U^{\prime} R_z \ket{0}
		= U^{\prime} \ket{0}
		\label{eq:prep}
\end{equation}
since $z$-rotations do not change $\ket{0}$. Similarly, operations immediately
prior to detection can be performed up to an $R_z$ without effecting the
probability to be in $\ket{0}$ or $\ket{1}$.

Prior to performing an MS gate, the stored unitary of all targeted ions must be
physically undone as arbitrary single-qubit operations do not generally commute
with the MS gate. Additionally, the relative frames of targeted ions must
be rotated to match the MS gate beams. It would seem we are forced to perform
exact single-qubit unitary operations around all MS gates. However, these
requirements can be softened. If we diagonalize the MS interaction with
appropriately rotated Hadamard gates on each ion, $R_z$ operations will commute and
can be tracked to the other side. This is the origin of the ``gate
optimization'' additions to \refsubfig{fig:CNOTseq}{a} and results in only two $\Rphi$
prior to the MS gate and three after it. Exact unitaries would require
six to eight $\Rphi$ operations.

In summary, our compiler stores a unitary history for operations on each ion.
This enables us to track crosstalk corrections, unperformed single-qubit gates,
and unperformed $R_z$ rotations. We are able to program algorithms directly from
a quantum circuit, and the compiler produces a compact set of operations which
limit the number of physical rotations actually performed.

\section{\label{app:echo}MS Echo}
We describe an ``echoing'' scheme that enables pairwise entanglement despite the
imperfect pairwise addressing of our MS gate beams. The intuitive motivation is
that flipping the state of an ion in the MS basis 
(i.e. $\ket{+}_x \leftrightarrow \ket{-}_x$)
halfway through the gate will
cause the net effect of the MS interaction on this ion to be identity.
In fact, we show that 
the scheme is exact (within the approximations outlined below)
and could be used to entangle any $m$ ions in an $N$ ion chain with $2^{N-m}$
single-qubit flips on the untargeted ions; 
these ``echo'' operations must be equally distributed throughout
the MS gate, requiring it to be performed in $2^{N-m}$ equal steps.

Following \cite{Sørensen2000}, we write the interaction picture Hamiltonian,
$\tilde{H}_{\mathrm{int}}$,
for two Raman beams with symmetric detuning $\delta$ from the carrier
transition, assuming identical intensity profiles for the red- and blue-detuned
beams. Assuming the detuning is close to one sideband frequency $\nu$, we
ignore all other modes as well as terms oscillating at $\nu+\delta$. In the
Lamb-Dicke approximation, $\tilde{H}_{\mathrm{int}}$ is given by
\begin{equation}
		\tilde{H}_{\mathrm{int}} \approx \sqrt{2} \Omega \eta \tilde{J}_x 
		\left[x \cos(\nu-\delta)t + p \sin(\nu-\delta)t \right] ,
		\label{eq:Hint-simp}
\end{equation}
where $\Omega$ and $\eta$ are the maximum Rabi rate and Lamb-Dicke
factor, respectively, for any ion in the chain,
$x$ is the position operator and $p$ the momentum operator (defined in
\cite{Sørensen2000}), $t$ is time,
and we have defined
\begin{equation}
		\tilde{J}_x = \sum_{i=1}^N c_i X_i
		\mbox{\quad and \quad} 
		c_i = \frac{1}{2} \frac{\Omega_i}{\Omega} \frac{\eta_i}{\eta} ,
		\label{eq:weightedJ}
\end{equation}
where $X_i$ is the Pauli $X$ operator acting on the $i$th ion.
The Rabi rate $\Omega_i$ and Lamb-Dicke factor $\eta_i$ are different for
each ion $i$ in the $N$-ion chain. \Eref{eq:Hint-simp} also neglects
off-resonant carrier transitions, which is a valid approximation 
provided $\Omega \ll \delta$ \cite{Sørensen2000}. This Hamiltonian is
a generalization of equation (6) in \cite{Sørensen2000}, 
and the results for the propagator $U$ follow directly with a modified
spin-dependence, $\tilde{J}_x$. If all ions had the same Rabi rate and Lamb-Dicke factor,
then all $c_i$ would be $\frac{1}{2}$ and $\tilde{J}_x$
would reduce to the total angular momentum operator $J_x$.

The authors of \cite{Sørensen2000} describe how the propagator resulting from a
Hamiltonian of the form \eref{eq:Hint-simp} causes eigenstates of $J_x$ to
traverse a circular path in phase space. For a propagation time 
$2\pi n /(\nu-\delta)$ with $n = 1,2,3\ldots$,
the phase space trajectory returns to the origin, and the ions are left
in the original motional state. At these times, the propagator
$U= \exp(-\rmi A \tilde{J}_x^2)$, where $A$ is proportional to the geometric
area enclosed by the loop in phase space.
In our general case, the proportionality factor
$\tilde{J}_x^2$ is now a sum over pairs of $X$ operators, weighted by the
relative intensity and Lamb-Dicke factor [$c_i$, defined in
\eref{eq:weightedJ}]. Simplifying, 
\begin{equation}
		\tilde{J}_x^2 = \sum_{i=1}^N \sum_{j=1}^N c_i c_j X_i X_j
		= \sum_{i=1}^N c_i^2 + 2\sum_{i=1}^N \sum_{j>i}^N c_i c_j X_i X_j
		\label{eq:jsquared}
\end{equation}
where the first term on the right-hand side is a global phase, independent of the internal
states of the ions; we ignore this global phase and redefine
\begin{equation}
		U \equiv \exp\left( -2 \rmi A \sum_{i=1}^N \sum_{j>i}^N c_i c_j X_i X_j\right) .
		\label{eq:propagator}
\end{equation}
We define an ``echo'' operation
\begin{equation}
		E_k = \exp\left(-\rmi \frac{\pi}{2} Y_k\right) = -\rmi Y_k
		\label{eq:flip}
\end{equation}
which swaps $x$-eigenstates of the $k$th ion through a $\pi$ rotation about the
$y$-axis. Unitary transformations of the propagator are built from the $E_k$:
\begin{eqnarray}
	U_{k_1 k_2 \cdots k_L} 
	&= \left(\prod_{l=1}^L E_{k_l}^{\dagger} \right) U \left(\prod_{l=1}^L E_{k_l} \right) \\
	&= \left(\prod_l E_{k_l}^{\dagger} \right) 
	\exp\left( -2 \rmi A \sum_{i=1}^N \sum_{j>i}^N c_i c_j X_i X_j\right)
		\left(\prod_l E_{k_l} \right) \\
	&= \exp\left[-2\rmi A \sum_{i=1}^N \sum_{j>i}^N c_i c_j 
		\left( \prod_l E_{k_l}^{\dagger}\right) 
		X_i X_j \left( \prod_l E_{k_l} \right) \right] \label{eq:exp}\\
	&= \exp\left\{-2\rmi A \sum_{i=1}^N \sum_{j>i}^N c_i c_j 
		\left[ \prod_l (-1)^{\delta_{k_l i} + \delta_{k_l j}} \right] X_i X_j \right\}
	\label{eq:transU}
\end{eqnarray}
where \eref{eq:exp} can be shown by expanding the operator exponential with a
Taylor series, and \eref{eq:transU} uses $X_i Y_k = (-1)^{\delta_{ik}} Y_k X_i$.

We assert that a subspace $D$ within an $N$-ion chain can be decoupled so
that only the ions not in $D$ participate in the net MS interaction through
products of these $U_{k_1\cdots}$ with $k_l \in D$. For example, the following
sequence will decouple the $k$th ion in a $N$-ion chain
\begin{equation}
		U_k U = \exp\left\{-2 \rmi A \sum_{i=1}^N \sum_{j>i}^N c_i c_j 
		\left[1+(-1)^{\delta_{k i} + \delta_{k j}} \right] X_i X_j \right\} .
		\label{eq:echo1of3}
\end{equation}
Since $j\ne i$, $k$ will will only equal $i$ or $j$ or neither. Therefore, if
$i$ or $j$ is equal to $k$, the term in \eref{eq:echo1of3} in square braces
vanishes; conversely, the terms where $k \ne i,j$ survive, exactly as desired
for an MS gate on all ions but the $k$th.

Generally, a sequence to decouple an $d$-qubit subspace $D$ can be constructed
from the product of the $U_{k_1 k_2 \cdots}$ where the ion labels $k_l$ are all
combinations of 0 to $d$ indices in $D$. \Tref{tab:decouple} summarizes possible
decoupling sequences up to $d=4$. The number of such operators is given by a sum
over binomial coefficients:
\begin{equation}
		\sum_{i=0}^d {{d}\choose{i}} = 2^d .
		\label{eq:echoCount}
\end{equation}
For an MS gate on $m$ ions out of an $N$ ion chain there are $d=N-m$ ions in
$D$, requiring the MS gate be performed in $2^{N-m}$ equal steps as initially
asserted. Finally, we note that since the transformed $U$ operators all
commute, they can always be rearranged to give a decoupling sequence with a
single $E_k$ between each MS interaction. Generally, this implies the ability to
optimize the echo sequence for any single-qubit control scheme.

\Table{\label{tab:decouple}Sequences to decouple up to $d=4$ ions. The
transformed $U$ operators all commute. In practice, there may be an optimal
order in which to perform these operations.}
	\br
	$d$ & Possible sequence \\
	\mr
	1 & $U_k U$ \\
	2 & $U_{kl} U_k U_l U$ \\
	3 & $U_{klm} U_{kl} U_{km} U_{lm} U_k U_l U_m U$ \\
	4 & $U_{klmn} U_{klm} U_{kln} U_{kmn} U_{lmn} 
	U_{kl} U_{km} U_{kn} U_{lm} U_{ln} U_k U_l U_m U_n U$ \\
	\br
\endTable

In \sref{sec:2of3}, we demonstrate the most rudimentary echoing scheme. By
decoupling one ion within a three-ion chain, we generate
pairwise entanglement despite imperfect pairwise addressing. Within a longer
chain, we note one could shorten the required echo sequence by neglecting to
echo ions which have a near-zero interaction ($c_i$) with the gate
beams.

While this echoing technique can decouple any arbitrarily connected subset of
ions, we have only considered a single vibrational mode under the assumption of
near-detuning. Spectator modes, which tend to lead to residual motional
entanglement, may be a key error source. In the future, we anticipate using
modulated sequences, like those described in \cite{Choi2014}, to alleviate this
issue. Such sequences could also be interleaved with echo pulses to arbitrarily
decouple selected ions with high fidelity.

\section*{References} 
\bibliographystyle{iopart-num}
\bibliography{refs}

\end{document}